\documentclass[%
  prab,
 twocolumn,
 amsmath,amssymb,
 reprint,%
 superscriptaddress,
 nofootinbib,
]{revtex4-2}

\usepackage{graphicx}
\usepackage{dcolumn}
\usepackage{bm}
\usepackage{color,soul}
\usepackage{cleveref}
\usepackage{siunitx}
\usepackage{physics}
\draft 

\begin{document}

\preprint{HIT-11-2021}

\title{Interpretation of the horizontal beam response near the third integer resonance} 

\author{E.C.~Cort\'es García}
\email[]{edgar.cristopher.cortes.garcia@desy.de}
\affiliation{Deutsches Elektronen-Synchrotron DESY, Notkestraße 85, 22607 Hamburg, Germany}
\affiliation{GSI Helmholtz Centre for Heavy Ion Rearch, Planckstraße 1, 64291 Darmstadt, Germany}

\author{P.\ Niedermayer}
\affiliation{GSI Helmholtz Centre for Heavy Ion Rearch, Planckstraße 1, 64291 Darmstadt, Germany}

\author{R.\ Singh}
\affiliation{GSI Helmholtz Centre for Heavy Ion Rearch, Planckstraße 1, 64291 Darmstadt, Germany}

\author{R.\ Taylor}
\affiliation{CERN, Espl. des Particules 1, 1211 Meyrin, Switzerland}
\affiliation{Imperial College London, Exhibition Rd., South Kensington, London SW7 2BX, United Kingdom}

\author{E.~Feldmeier}
\affiliation{Heidelberg Ion-Beam Therapy Center, Im Neuenheimer Feld 450, 69120 Heidelberg, Germany}

\author{M.~Hun}
\affiliation{Heidelberg Ion-Beam Therapy Center, Im Neuenheimer Feld 450, 69120 Heidelberg, Germany}

\author{E.\ Benedetto}
\affiliation{SEEIIST Association, Rue de Battoirs 7, 1205 Geneva, Switzerland}

\author{T.~Haberer}
\affiliation{Heidelberg Ion-Beam Therapy Center, Im Neuenheimer Feld 450, 69120 Heidelberg, Germany}

\date{\today}

\begin{abstract}
  The beam response to an external periodic excitation delivers relevant information about the optics, tune distribution and stability of a circulating beam in a storage ring.\ In this contribution the horizontal beam response to the excitation (transfer function) under conditions typical for slow extraction is presented for a coasting beam.\ The resulting spectrum exhibits a splitting behaviour.\ The single particle dynamics is discussed and an interpretation based on simulation results is presented.
\end{abstract}

\maketitle
\section{Introduction}
Slow extraction from synchrotrons and storage rings is a well established method to deliver a variety of beams to experiments and patients.
In current versions of the method, the slow extraction is prepared by moving the horizontal working point near a third integer resonance while simultaneously reducing the dynamic aperture (DA) by driving the resonance with dedicated sextupoles.
The effective horizontal dynamics under such conditions can be described by the Kobayashi Hamiltonian~\cite{KobaHamiltonian}.
The beam is then extracted either by slowly reducing the DA further until it reaches zero, or by slowly increasing the beam emittance beyond the (constant) DA.
Different variations of this machine setup are in use~\cite{medsyncPullia, NodaPTReqs, SXTransitTime_RSinghPRA2020}. In this contribution, the so called Radio Frequency Knock Out (RF-KO) extraction is considered, consisting of blowing-up the emittance through an external transverse excitation while keeping the DA constant~\cite{TOMIZAWA1993}.
\\
In RF-KO extraction, the emittance blow-up is achieved by applying an external RF electromagnetic field excitation.
The amplitude of the betatron oscillations is thereby controlled and eventually increased beyond the DA where particles become unstable and are pushed into the extraction channel.
Several studies have been performed in order to engineer an optimal frequency spectrum for the RF excitation signal~\cite{RFKOSX_NodaNIMA1996, NodaExtDiffRegion, NodaDualFM, Krantz, YamaguchiMultiMode, Shiokawa2021}; these studies have heavily relied on signals of extracted particle counts for trial-and-error optimization of the RF-signal spectrum.
In this contribution we take a step back and focus exclusively on the amplitude and phase response of the stored beam to an external sinusoidal excitation by measuring the beam transfer function (BTF) under extraction conditions.
Here the sinusoidal stimulus replaces the usual RF excitation signal, but the rest of the machine setup remains the same as for RF-KO extraction.\\
\\
Studies of BTFs close to driven resonances are scarcely found in the literature.
The information gained by the BTF measurements is of utter importance, since it has a significant application in the context of engineering an optimal waveform for the slow extraction~\cite{HIT2022, Niedermayer_2024}, where fixed tone excitations have also been proposed~\cite{Niedermayer_2024}.\\
\\
Our contribution is organized as follows: In Section~\ref{sec:theory} the particle dynamics under slow extraction conditions is reviewed together with the theoretical background of the extraction, for single and multi-particle dynamics.
In Section~\ref{sec:experiment} and \ref{sec:simulation}, measurements and simulations of the beam response of a coasting beam are presented respectively.
We culminate with a summary and a discussion of the results in Section~\ref{sec:summAndDiscussion}.

\section{Theory}
\label{sec:theory}
\subsection{Single particle dynamics}
The status-quo of the study of non-linear dynamics relies on the Lie-algebraic formulation of transfer maps induced by Hamiltonians of different particle accelerator elements~\cite{Dragt,ChaoSpecialTopics}.
The workhorse for the engineering and optimization of slow extraction via the third integer resonance is the Kobayashi Hamiltonian~\cite{KobaHamiltonian}.
It is an approximate description of the particle dynamics when the tune~$q_x$ is close to a third order resonance~$q_\text{res}$ driven by sextupole magnets.
Comprehensive derivations and corollaries of this dynamics can be found in~\cite{PIMMS}.

\begin{figure}
  \includegraphics[width=\linewidth]{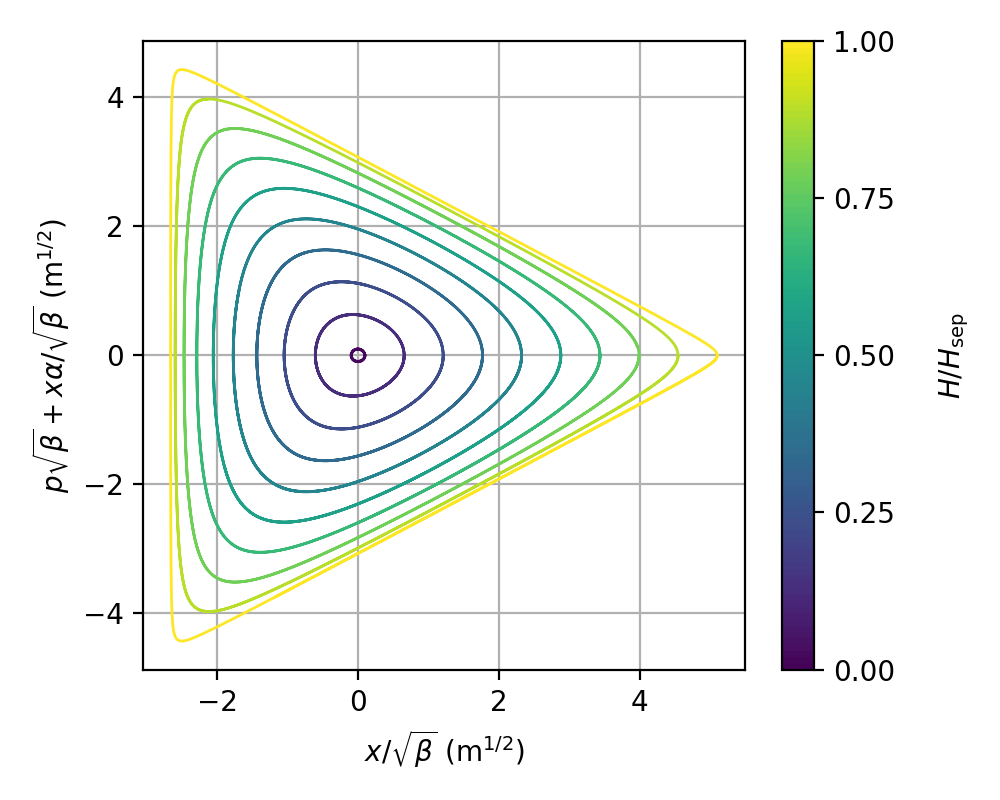}  \caption{\label{KobayashiPhaseSpace} Equipotential lines in the stable region of normalized phase-space induced by the Kobayashi Hamiltonian.\ The value of the Hamiltonian at the separatrix edge reads $H_{\mathrm{sep}} = [4\pi(q_{x,0} - q_{\text{res}})]^3/S^2$ \cite{PIMMS}.}
\end{figure}

We identify the Kobayashi Hamiltonian~$H$ using the well established action-angle variables $(J_x, \phi_x)$ which relate to the horizontal space coordinate~$x$ as $x = \sqrt{2\beta_x J_x} \sin \phi_x$~\cite{ChaoSpecialTopics,PIMMS,ThirdIntegerExp}:
\begin{align}
  H = h_0 + h_1
  \label{eq:KobayashiHamiltonian}
\end{align}
with
\begin{align}
  h_0 &= - 6\pi (q_{x,0} - q_{\text{res}})J_x\ =: - \delta J_x , \\
  h_1 &= - \frac{S}{\sqrt{2}} J_x^{3/2} \sin{3\phi_x},
\end{align}
where $S$ can be interpreted as the amplitude of the resonance driving term excited by an effective single virtual sextupole.\
The explicit expression for the computation of the amplitude $S$ of the resonance driving term given a set of sextupoles distributed in the lattice can be found in Appendix \ref{sec:ResonanceDrivingTerm}.\
The linear (zero-amplitude) tune in absence of the nonlinear sextupole fields is denoted as $q_{x,0}$, and $q_\text{res}$ is the closest third integer resonance tune ($3q_\text{res} \in \mathbb{N}$).
Figure~\ref{KobayashiPhaseSpace} depicts the equipotential lines of $H$ in phase-space in the range where stable motion is possible.

\subsubsection{Amplitude-phase-detuning}
Since $H$ describes the effective motion of a particle every three consecutive turns, using the Lie-algebraic treatment~\cite{ChaoSpecialTopics, WHerrCAS}, the amplitude-phase dependent tune over three consecutive turns follows as:
\begin{equation}
  q_x 
  = q_{x,0} + q_{x,1}
  = \frac{1}{6\pi}\frac{\partial H}{\partial J_x} + q_{\text{res}}
  \label{eq:TuneFromHami}
\end{equation}
where $q_{x,0}$ is the unperturbed tune given by the linear theory and 
\begin{align}
  q_{x, 1} =q_{x, 1}(J_x, \phi_x) = \frac{3S}{\sqrt{2^5}\pi}J_x^{1/2}\sin{3\phi_x}
  \label{eq:KobaDetuning}
\end{align}
is the amplitude-phase dependent tune shift due to the nonlinear dynamics averaged over three turns.
It is common to apply averaging techniques over $\phi_x$ to Eq.~\ref{eq:KobaDetuning} and intuitively, the contribution from the phase-amplitude detuning for a nearly constant action $J_x$ should vanish.
\\

\begin{figure}[b!]
  \includegraphics[width=\linewidth]{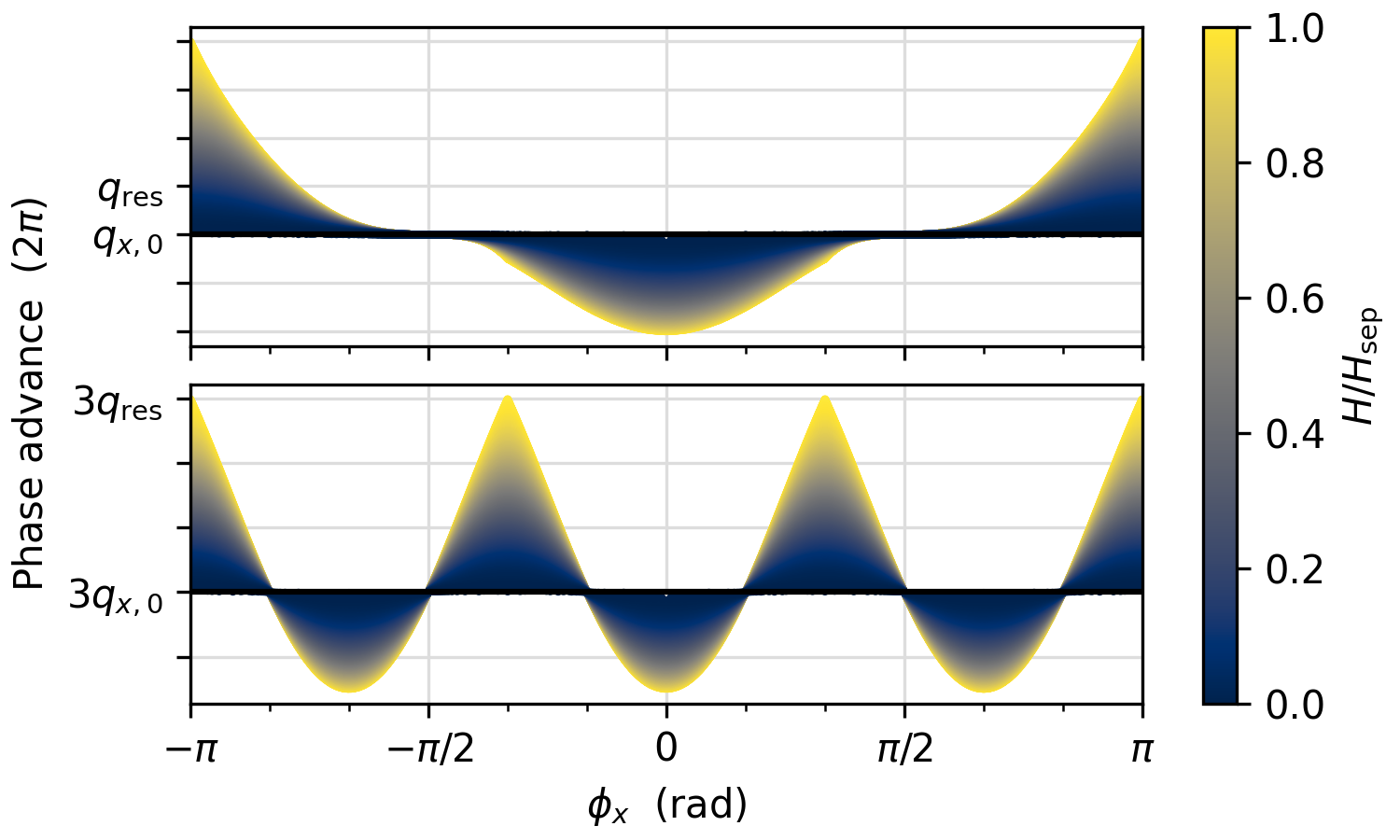}
  \caption{Single (top) and three turn phase advance (bottom) as function of angle $\phi_x$, coloured by the value of the Kobayashi Hamiltonian $H$.}
  \label{amplitude-phase-detuning}
\end{figure}

However, near a resonance the phase space is deformed as the Kobayashi Hamiltonian clearly shows (Fig.~\ref{KobayashiPhaseSpace}), and the action $J_x$ is \emph{not} an invariant of motion~\cite{Bazzani262179}.
Rather, it is a function of the angle $J_x = J_x(\phi_x)$ and as such modulated even without the effect of an external excitation.
This is an important aspect: the phase advance of a single particle is not constant in time, but varies as a function of the angle as the particle revolves in phase space. The resulting phase-advance modulation is illustrated in Fig.~\ref{amplitude-phase-detuning}.
As discussed later, this aspect determines how the particle responds to a sinusoidal excitation at a certain frequency over time.

\begin{figure}[h]
    \centering\includegraphics[width=\linewidth]{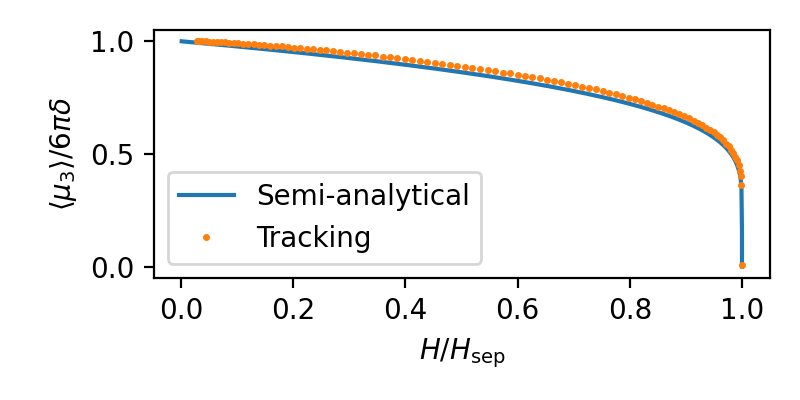}
    \caption{Average three-turn phase advance as a function of the Hamiltonian.\ The solid blue line shows the value of Eq. \ref{eq:PNiedermayerAveDetuning}, while the orange dots show the result of single particle tracking of a FODO cell with a sextupole (S=-30) with $\delta = - 2.5 \cdot 10^{-3}$ for 3300 turns.}
    \label{fig:AverageDetuningVsEnergy}
\end{figure}

Moreover, the single particle detuning can be better understood by investigating its long term average behaviour \cite{Niedemayer_IntDoc}.\ The time average of a system's quantity $g$ is given by $\langle g \rangle = \int_{t_1}^{t_2} g \dd t / \int_{t_1}^{t_2} \dd t$.
Thus, the average three-turn phase advance $\langle \mu_3 \rangle$ yields \cite{Niedemayer_IntDoc}
\begin{align}
    \frac{\langle \mu_3 \rangle}{6\pi\delta}= \left(\frac{1}{2\pi}\int_{0}^{2\pi} \frac{\dd \phi_x}{1-\sqrt{2\tilde{J}}\cos{3\phi_x}}\right)^{-1}
    \label{eq:PNiedermayerAveDetuning}
\end{align}
where the normalized action $\tilde{J}$
\begin{widetext}
\begin{align}
  \tilde{J} = &\begin{cases}
    \tilde{H}/6 & \mathrm{if} \cos{3\phi_x} = 0 \\
    \frac{6 + R^{1/3}(\sqrt{3}i-1) + R^{-1/3}(\sqrt{3}i+1)(8\tilde{H}\cos^2{(3\phi_x)}- 9)}{16\cos^{2}{(3\phi_x)}} & \mathrm{otherwise}
  \end{cases},\\
  R = &27 - 36\tilde{H}\cos^{2}{(3\phi_x)} + 4\tilde{H}^{2}\cos^{3}{(3\phi_x)}\left( 2\cos{(3\phi_x) - \sqrt{2\cos{(6\phi_x)} +2 - 4/\tilde{H}}}\right)\nonumber
  \label{eq:PNiActionAsFctOfPhaseFormula}
\end{align}
\end{widetext}
and the normalized effective energy 
\begin{align}
    \tilde{H} = \frac{H}{H_{\text{sep}}} = \frac{H}{(4 \pi \delta)^3/S^2}
\end{align}
were introduced.\
Note that the r.h.s of Eq.~\ref{eq:PNiedermayerAveDetuning} is unity for $\tilde{J}\rightarrow 0$ and thus the average three-turn phase advance is consistent with the linear theory.\ When the energy of a particle approaches the limit of $H \rightarrow H_{\text{sep}}$, the denominator inside the integral diverges and therefore the average three turn phase advance freezes, i.e. the particle hits the resonance $\langle \mu_3 \rangle \rightarrow 0$.\ The behaviour of the average detuning over many turns is illustrated in Fig \ref{fig:AverageDetuningVsEnergy}.\ It is clearly visible that the average three-turn phase advance detunes towards the nearest third-order resonance.\ The effect is more drastic when the energy of the particle is closer to the edge of the separatrix. 

\subsubsection{External excitation}
A couple of archetypes of externally driven non-linear oscillator systems have been
extensively studied, e.g. the Duffing equation, the forced Van-Der Pol oscillator
\cite{Hayashi2014} or Chirikov's standard map \cite{BorisChirikov1979}, to mention a few
of the most prominent.\ In the context of particle accelerators considerable progress on the
theory of externally driven non-linear systems has been done
\cite{AdiabaticTrapping_BCG_PRAB2022, HiraiwaPRAB2021}, where the introduction of external
sinusoidal excitations provide useful schemes to either trap in non-linear resonance islands
fractions of the beam or to diffuse the beam in an RF abort case.\
Following the Lie-algebraic treatment, the effective time-dependent one-turn Hamiltonian of our system reads \cite{ChaoSpecialTopics}
\begin{align}
    H(n) &=  - \mu_xJ_x - \frac{h_1}{3} + A_x J_x^{1/2} \sin{\phi_x}\sin{(\Omega n)}\nonumber\\  
    &- \mu_x\frac{S}{\sqrt{2}}J_x^{3/2}\left(\frac{1}{3}\cos{3\phi_x}+\cot{\frac{\mu_x}{2}}\sin{\phi_x} + \cos{\phi_x}\right)
\label{eq:ExtDrivenOneTurn}
\end{align}
where $\mu_x = 2\pi q_x$ is the one-turn phase advance, $A_x$ is the normalized strength of the external periodic excitation, $\Omega$ is its frequency  and $n$ is the turn number.\ In this case the system does not resemble any of the archetypes.\ 

\subsection{Collective dynamics: Beam response}
The study of the collective response of an ensemble of oscillators under the influence of an external weak force $ f_x(t) = \epsilon \cos(\Omega t)$ can be found in literature for the analysis of beam instabilities and study of tune spectra (see e.g.\ \cite{TransverseBTF_SinghPRAB2013, BeamBeamBTF_GoergenPRAB2015}).\ 
The amplitude of the beam centroid oscillation $\langle x  \rangle$ as a function of the driving frequency (approx.) reads \cite{ChaosBook,Book_NgKY2005}
\begin{align}
    \langle x  \rangle(t) &= \frac{\epsilon}{2\mu_x} \left[\cos\Omega t\ \mathcal{P.V.} \int \dd \mu \frac{\rho(\mu)}{\Omega - \mu} + \pi\rho(\Omega) \sin\Omega t\right]
    \label{eq:SimpleBeamResponse}
\end{align}
where $\epsilon$ is the normalized strength of the force, $\rho(\mu)$ is the distribution function of the frequencies of the oscillator ensemble, $\mu_x$ is the linear betatron frequency (assumed to be the mean value of the distribution) and $\mathcal{P.V.}$ denotes the principal value of the integral.\ Eq.~\ref{eq:SimpleBeamResponse} is valid when $\rho(\mu)$ only peaks once and the transversal motion of the oscillators do not influence $\rho(\mu)$ itself.\ 
For the case where the oscillators in the ensemble experience a linear detuning as a function of their action $J_x$ a similar result is known \cite{BergAndRuggiero, LHC-Octupoles-2020}.\\
Hence an analytical expression is known for the beam transfer function (BTF) $T(\Omega)$
\begin{align}
    T(\Omega) = f(\Omega) + ig(\Omega)
\end{align}
with the conventional definitions 
\begin{align*}
    f(\Omega) = \mathcal{P.V.} \int \dd \mu \frac{\rho(\mu)}{\Omega - \mu}, && g(\Omega) = \pi \rho(\Omega).
\end{align*}
Note that the beam centroid oscillation amplitude (Eq.\ref{eq:SimpleBeamResponse}) is directly related to the BTF by
\begin{align}
    \langle x  \rangle(t) = \Re\left[\frac{\epsilon}{2\mu_x}\exp(-i\Omega t) T(\Omega) \right].
\end{align}
For a system where the non-linear dynamics near a resonance are prevalent, there is no known analytical or semi-analytical expression for the BTF.\ Therefore we investigate experimentally the beam centroid oscillation amplitude to an external sinusoidal excitation under the aforementioned conditions.\ In the following we will refer to this as the beam response.

\section{Beam response measurement}
\label{sec:experiment}
Measurements of the beam response under resonant slow extraction conditions were carried out at the Heidelberg Ion-beam Therapy Centre (HIT) in 2022. 
In this section the measurement setup and experimental results are presented.

\subsection{Experimental setup} 
\subsubsection{Heidelberg Ion-Therapy Centre Synchrotron}
\begin{figure}[h!]\includegraphics[width=\linewidth]{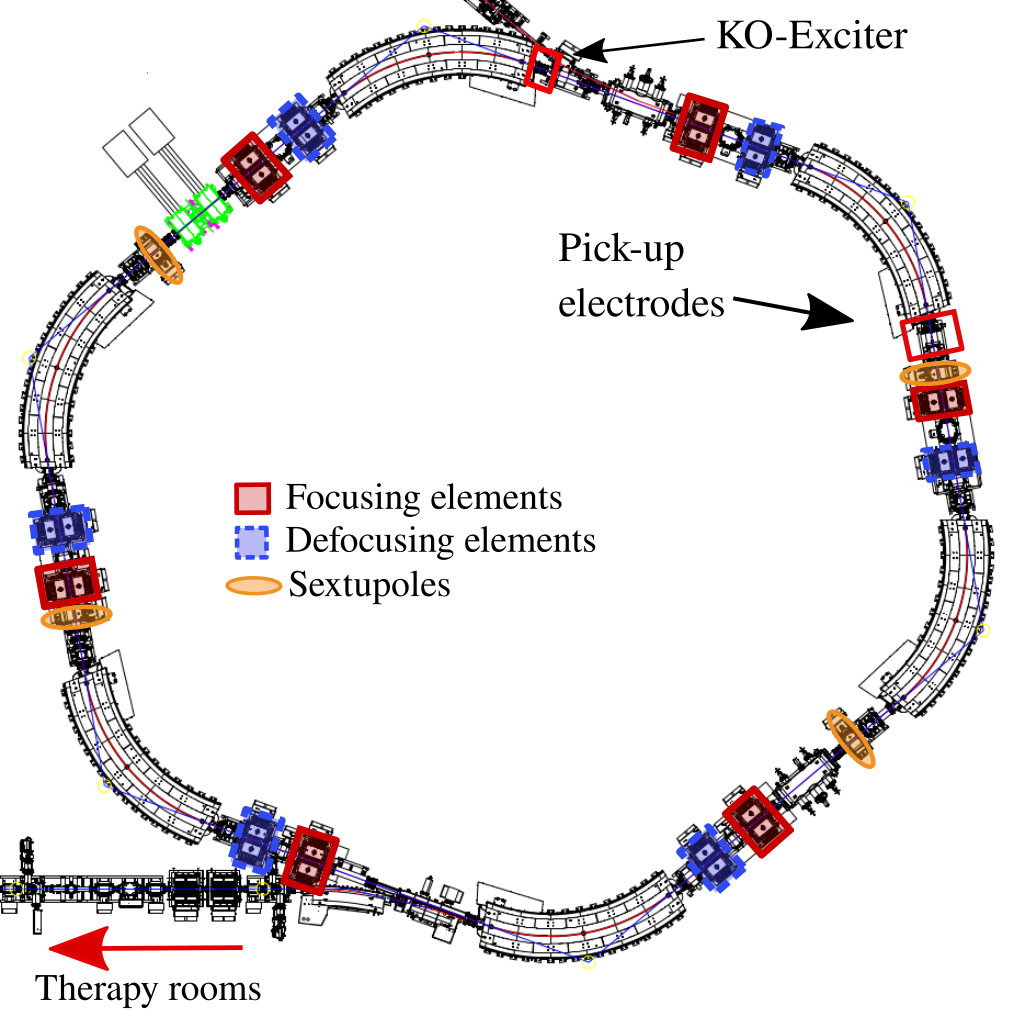}
  \caption{\label{fig:HIT}The Heidelberg Ion-beam Therapy Center's synchrotron. The position of the (de)focusing quadrupoles are indicated with (blue)red boxes. The position of the sextupoles are indicated with an orange ellipse. The injection channel can be seen in the top middle, whereas the extraction to the therapy rooms is visible in lower left corner.}
\end{figure}
\begin{figure}
    \centering
    \includegraphics[width=\linewidth]{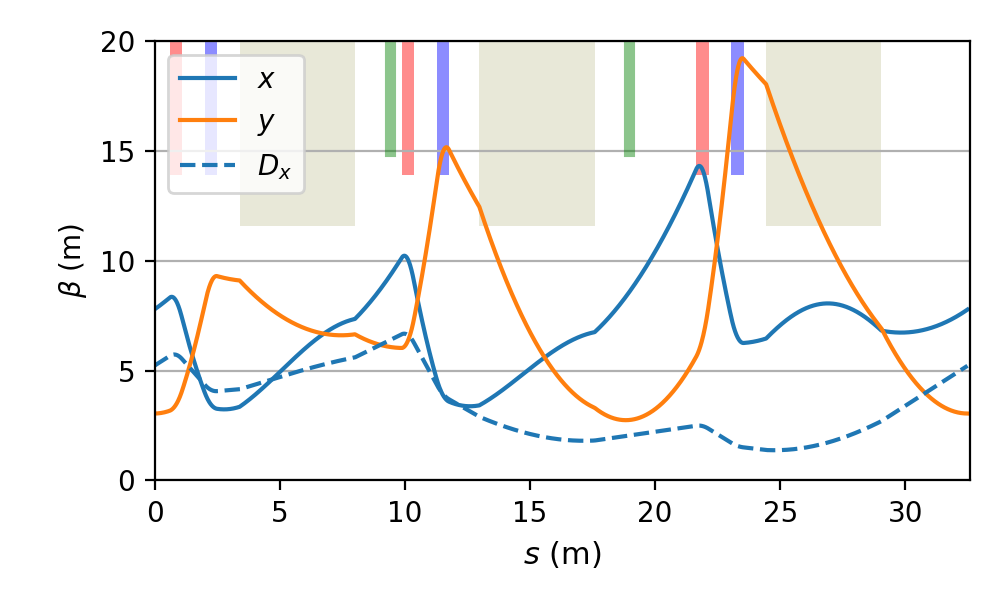}
    \caption{Optical functions of the HIT's synchrotron super-cell at extraction conditions.\ The position of the (de)focusing elements are marked with a red(blue) box on  the top. The green marks indicate the position of the sextupoles.\ The dipoles are illustrated with a wide box.}
    \label{fig:optFuncsHIT}
\end{figure}

The Heidelberg Ion-Beam Therapy Centre (HIT) is a medical accelerator facility dedicated to the treatment of cancer tumours in patients with light ions ($\text{p}^{+}, \text{He}^{2+}, \text{C}^{6+}$ and $\text{O}^{6+}$) using the intensity controlled raster scan \cite{HITProposal,HabererScanSystem}.\ 
The HIT synchrotron layout is illustrated in Fig.~\ref{fig:HIT}.\ It is a compact machine with a circumference of \SI{64.96}{\meter}.\ 
The ring is composed of two super-cells each with three quadrupole doublet sections.\ 
Each super-cell is equipped with two families of sextupole magnets, one optimally placed for chromaticity correction and the other one for the excitation of the resonance driving term.\ 
The optical functions of the synchrotron at extraction conditions are depicted in Fig.~\ref{fig:optFuncsHIT}. 

The measurements are carried out with a Carbon-ion beam $^{12}\text{C}^{6+}$ with $E_{\text{kin}} = \SI{124.25}{MeV/nucleon}$.
There was a dedicated measurement campaign to characterize the linear optics of the machine and the beam properties relevant for the study.
Unfortunately, the beam diagnostics at HIT do not enable the direct measurement of the transverse emittance of the coasting beam, hence this has to be kept as a free parameter. 
The measured machine and beam parameters are listed in Table~\ref{tab:ExpSetup}.
\begin{table}
\caption{\label{tab:ExpSetup} Machine and beam parameters for the beam response measurements.\ The model values come from a calculation from MADX \cite{MADX}.}
\begin{ruledtabular}
  \begin{tabular}{ c  c  c }
    Parameter & Value & Model \\
    \colrule
    $q_x$ / $q_y$ & 1.679(1) / 1.720(1) & 1.67895 / 1.755\\
    $\xi_x / \xi_y$ & -1.70(2)/-1.63(2) & -1.78/-1.35\\
    $\eta$  & 0.46(3)  & 0.47657\\
    \hline\hline
    $\Delta p/p$ (RMS) & 0.5$\times 10^{-3}$ & \\
    $f_{\text{rev}}$ & \SI{2.1721}{\mega\hertz} & \\
    $N_P$ & 8$\times$10$^{7}$ particles
  \end{tabular}
\end{ruledtabular}
\end{table}

\subsubsection{Beam response measurement setup}

\begin{figure}[!h]
  \includegraphics[width=0.75\linewidth]{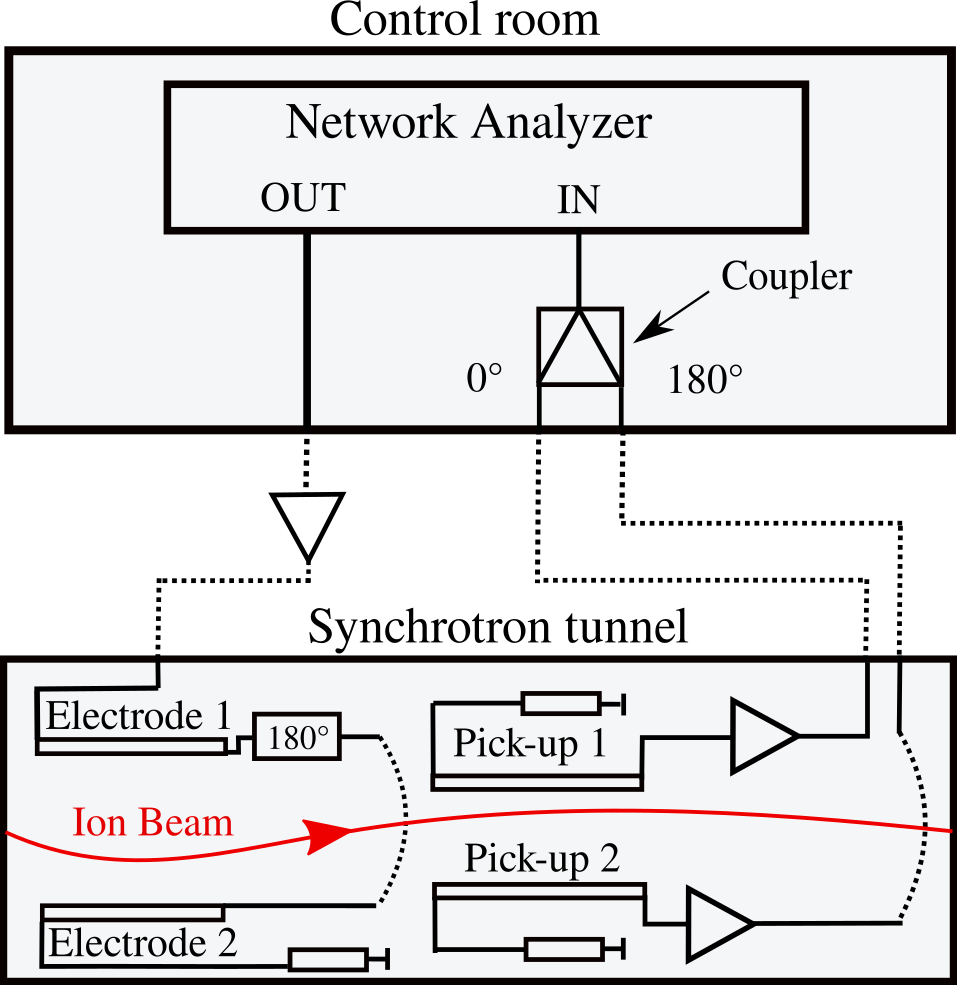}
  \caption{\label{fig:BTFSetup}Schematic of the experimental setup at HIT for the measurement of the beam response to an external excitation.}
\end{figure}

The beam response to an external excitation is measured with a vector network analyzer (VNA).
The stimulus signal is connected to the amplifier of the regular KO-Exciter, providing transverse kicks to the beam.
These sinusoidal kicks are of the form $\Delta x' = k_0l \sin(2\pi f_\mathrm{ex} t + \varphi_\mathrm{ex})$ with the stimulus frequency $f_\mathrm{ex} = q_\mathrm{ex} f_\mathrm{rev}$ and phase $\varphi_\mathrm{ex}$.
The estimated transverse kick amplitude due to a given input stimulus power can be found in Table~\ref{tab:StimulusPowerToKick}.\
The stimulus leads finally to transverse oscillations of the beam centroid, which are recorded with regular capacitive pick-up electrodes in difference mode.\
This provides the amplitude and phase dependent signal to the VNA, which calculates the system response at the stimulus frequency.\
The setup is depicted in Fig.~\ref{fig:BTFSetup}.\
Each measurement was carried out in a single machine cycle, where the stimulus frequency was swept over the course of \SI{10}{\second}.\
All the beam response measurements presented here were performed with a coasting (unbunched) beam.\

\begin{table}
    \centering
    \caption{Estimated maximum kick for a given input stimulus power with a $^{12}\text{C}^{6+}$ Carbon-ion beam with $E_{\text{kin}}=$~\SI{124.25}{\mega e\volt/nucleon}. The values were estimated with a FEM simulation of the HIT KO-Electrodes.}
    \begin{tabular}{c c}
    \hline\hline
    Input power & Kick\\
    \hline
        0 dBm &  \SI{1.54}{\micro\radian} \\
      -10 dBm &  \SI{486}{\nano\radian}\\
      -20 dBm &  \SI{154}{\nano\radian}\\
      -30 dBm &  \SI{48.7}{\nano\radian}\\
    \hline\hline
    \end{tabular}
    \label{tab:StimulusPowerToKick}
\end{table}

\subsection{Experimental results}
\begin{figure}
  \includegraphics[width=1.0\linewidth]{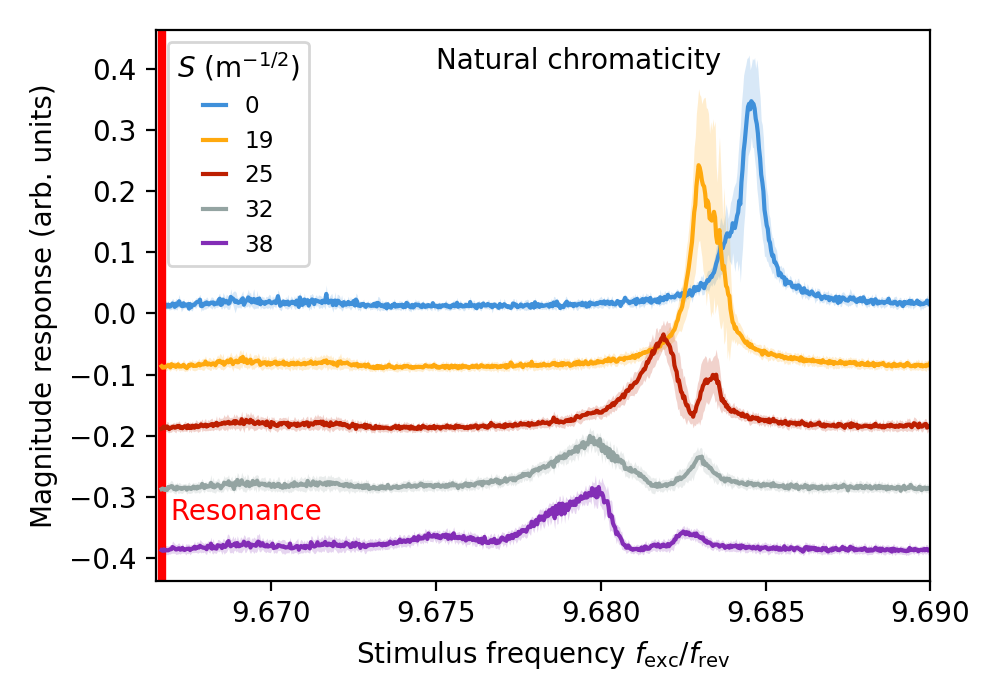}
  \includegraphics[width=1.0\linewidth]{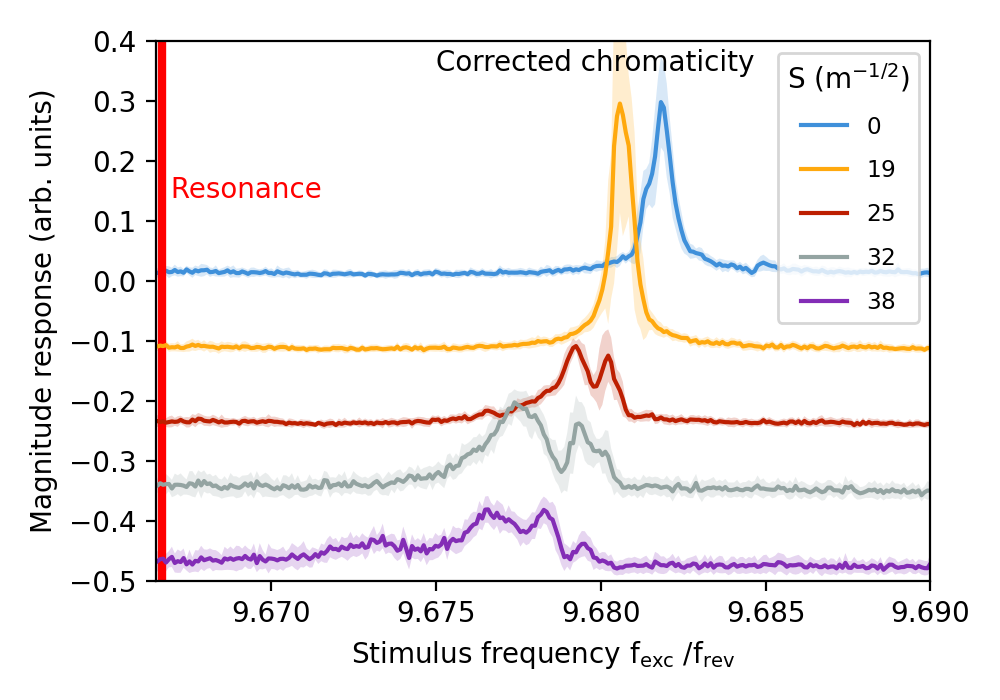}
  \caption{
    Beam response measurement as function of sextupole strength at the 9$^\text{th}$ upper harmonic of the betatron resonance, with natural chromaticity (top) and fully compensated horizontal chromaticity (bottom).
    The stimulus frequency is scanned from the resonance (red line) to higher frequencies
    with a strength of \SI{-30}{dBm} ($k_0l = \SI{48.7}{\nano\radian}$).
    Each graph composes the average and standard deviation over 25 cycles.
    The data was intentionally staggered in magnitude response and for $S \in [32,38]$ it was multiplied by a factor of two for the corrected chromaticity case for better visualization.
  }
  \label{fig:BeamResponse9thUpperHarm}
\end{figure}

The beam response was measured in the vicinity of the upper betatron sideband of the 9th revolution harmonic and the lower sideband of the 8th harmonic.\
The integrated strengths of the sextupole magnets $k^\prime_SL_S$ were varied in order to achieve a variation in the effective sextupole strength $S$ (see Eq.~\ref{eq:VirtSextupoleMag}).

\begin{figure}
  \includegraphics[width=1.0\linewidth]{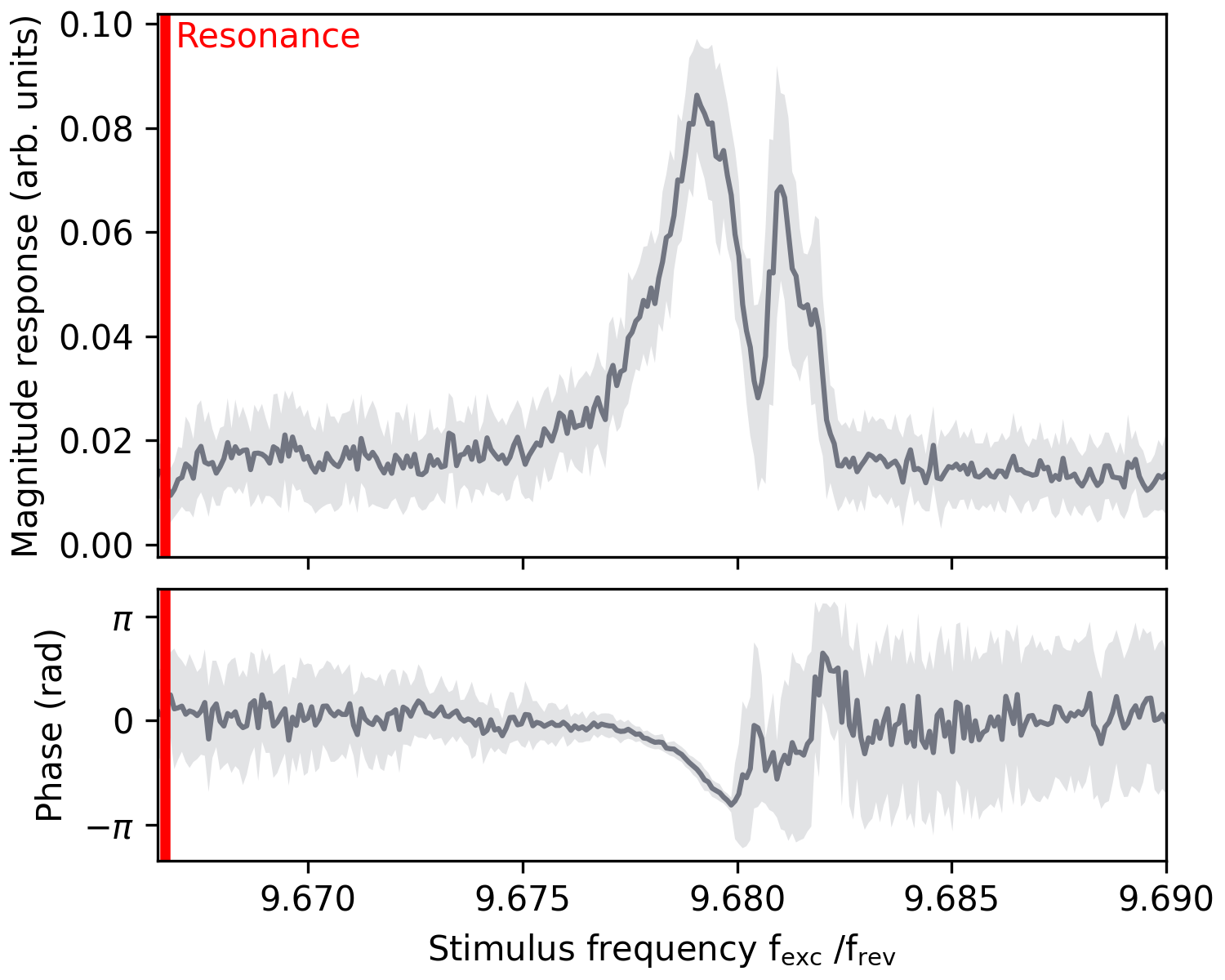}
  \caption{
    Average beam response (solid curve) measurement of 25 machine cycles with $S =$~\SI{32}{m}$^{-1/2}$ from the sextupole strength scan with corrected chromaticity.\ The rms of the measurements is shown as the shaded area.
  }
  \label{fig:PhaseResponseOfSingleMeasurement}
\end{figure}

The results are illustrated in Fig.~\ref{fig:BeamResponse9thUpperHarm} for an excitation span near the upper band of the 9th harmonic of the betatron resonance. 
The data shown was intentionally displaced (staggered) for better visualization of the details.\ 
Note that for the corrected chromaticity case (lower panel) for $S \in [32,38]$ the magnitude response was multiplied by a factor of two.\
The phase response for these measurements overlap considerably due to noise and is therefore not presented in Fig.~\ref{fig:BeamResponse9thUpperHarm}.\ 
An example of the phase response with $S =$~\SI{32}{m}$^{-1/2}$ is shown in Fig.~\ref{fig:PhaseResponseOfSingleMeasurement}.\
The region where the magnitude response splits exhibits fluctuations in the phase.\
For $S \in [0, 19] \text{m}^{-1/2}$ the beam response is very similar.\ Both of the curves are in the regime where the transversal dynamics are dominated by $h_0$ from Eq.~\ref{eq:KobayashiHamiltonian}, the motion resembles a linear oscillator and the non-linearity still does not play an important role.
Once $S \geq$ \SI{25}{\meter}$^{-1/2}$ the beam response starts splitting and exhibits two predominant maxima or one characteristic dip, respectively.\ The higher the value of $S$ is set, the wider the gap between the two peaks becomes.\
It is also evident that the mean value of the beam response shifts towards the third order resonance.\
This behaviour comes both from the amplitude-phase detuning expected from the underlying dynamics described by Eq.~\ref{eq:PNiedermayerAveDetuning}, as well as from a  contribution from the closed orbit at the sextupoles.\
The estimated closed orbit rms is \SI{5}{\milli \meter} with a max/min excursion of \SI{8}{\milli \meter}/\SI{-15}{\milli \meter}.\

\begin{figure}
  \includegraphics[width=1\linewidth]{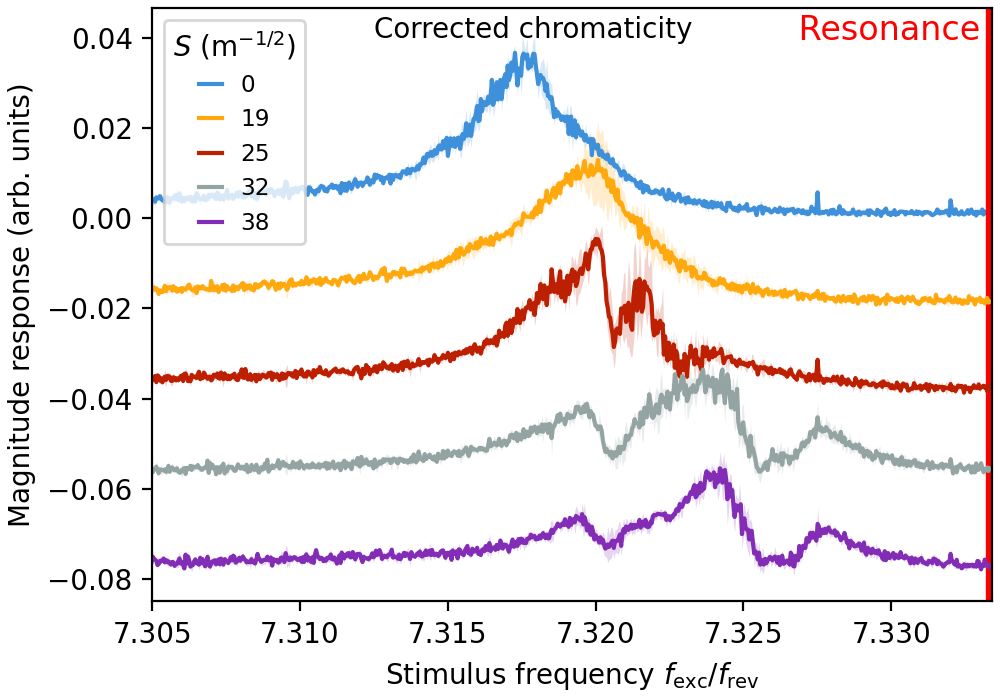}
  \caption{
    Beam response measurement as function of sextupole strength at the 8$^\text{th}$ lower harmonic of the betatron resonance, with fully compensated horizontal chromaticity.
    The stimulus frequency is scanned from lower frequencies towards the resonance (red line)
    with a strength of \SI{-10}{dBm} ($k_0l = \SI{486}{\nano\radian}$).
    Each graph composes the average and standard deviation over 3 cycles.
    The data was intentionally staggered for better visualization.\
  }
  \label{fig:BeamResponse8thLowerHarm}
\end{figure}

The same sextupole scan was performed and the beam response was measured at the lower 8th betratron harmonic.\
The results are shown in Fig.~\ref{fig:BeamResponse8thLowerHarm}.\
The measured phase response exhibited strong fluctuations as the 9th upper band.\ Thus the data is not shown.\
Up until  $S \in [0, 19] \text{m}^{-1/2}$, there is one single peak in the beam response, indicating that the motion is governed by the linear dynamics.\
Above this threshold the non-linearity induces a splitting in the measured beam response. In this case one can recognize more than two maxima or respectively at least one characteristic dip.\
The measurement of the beam response for the 8th lower harmonic and 9th higher harmonic were performed with the stimulus frequency brushing from the lower end to the higher end of the frequency span corresponding to each setting.\
This means that in the case of the 9th harmonic shown in Fig.~\ref{fig:BeamResponse9thUpperHarm}, the stimulus frequency went first through the 3rd order resonance (highlighted as a vertical line in red on the left), then through the betatron resonance where the beam sits and continued to the higher end of the span.\
On the contrary, for the scan near the 8th harmonic shown in Fig.~\ref{fig:BeamResponse8thLowerHarm}, the stimulus started with a frequency far from the resonance, then passed through the betatron resonance and ended at the third order resonance.\
This is an important aspect of the measurement that has to be considered, since the expected detuning due to the non-linear dynamics only shifts the single particle tune towards the resonance. In other words, the direction of the excitation sweep can lead to different dynamics.\

\begin{figure}
  \includegraphics[width=\linewidth]{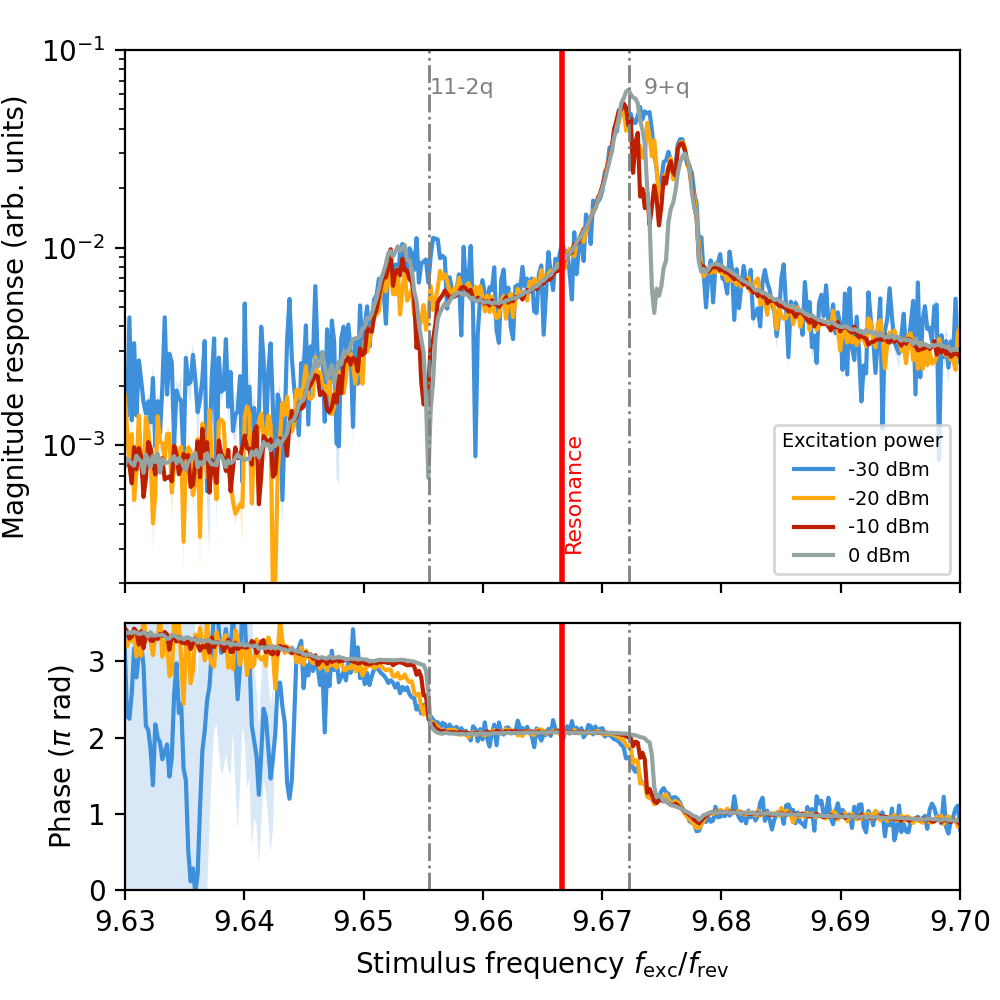}
  \caption{
  Beam response with a $S=$\SI{28}{m}$^{-1/2}$ for different input stimulus strengths.
  The dash-dotted lines indicate, which resonance the peaks correspond to.
  The double peak structure becomes stronger with increasing excitation power.}
 \label{fig:BeamResponseStimulusPowerScan}
\end{figure}

Finally, the dependency of the stimulus power of the beam response was measured near the 9th upper betatron resonance as well.\ The results are shown in Fig.~\ref{fig:BeamResponseStimulusPowerScan}, where the upper panel shows a broad scan in the stimulus frequency with the different curves showing the variation of the stimulus power (see Table~\ref{tab:StimulusPowerToKick} for the corresponding kick amplitude).\
The red line indicates the position of the nearest third order resonance.\
The grey dashed lines indicate the approximate expected positions of different betatron resonance bands.\
The lower panel shows the phase with regard to the stimulus. The phase jumps are clear when the frequency crosses the dip in the magnitude response.\ Note that the phase response data was intentionally unwrapped\footnote{This means that a sudden phase jump in adjacent phase differences are never greater than $\pi$ by adding $2\pi k$ for some integer $k$.\ This signal processing routine is available in the standard numpy library \cite{numpyLibrary}.}.\
By increasing the input power of the stimulus, which corresponds to an increasing strength of the kick amplitude that the beam experiences, the dip imprints deeper in the measured beam response.\

\section{Beam response simulation}
\label{sec:simulation}
The simulation of the beam response measurement is performed by launching a campaign of particle tracking computations to recover the horizontal beam oscillation amplitude.\
The particle tracking is performed with Xsuite \cite{XSuite}.\
The closed orbit corrector strengths are included in the model.\
A set with $N_p = 10^5$ particles is initialized, where each particle is described by an initial six-dimensional state vector $\vec{z}$.\ 
The transverse set of conjugate pairs $(x,p_x)$ and $(y,p_y)$ are initialized such that their probability density function is described by a stationary two-dimensional Gaussian distribution.\
Both transversal planes are assumed to be uncorrelated and the transverse emittance ratio is set to unity.\ The initial transversal emittance was set to $\varepsilon_{x,y} =$\SI{1}{\micro\meter \radian}.\
The longitudinal position coordinate is distributed uniformly in the ring to resemble the coasting beam, whereas the momentum spread is modelled with a Gaussian distribution with an rms value of $\sigma_{p} = 0.5\times 10^{-3}$, which is the value listed in Table~\ref{tab:ExpSetup}.\
After initialization, the particles were tracked for $10^{4}$ turns to let the distribution converge to a stationary distribution for a given value of $S$.\
Then the particles experienced the kicks from the external stimulus signal.

The stimulus from the measurement is modelled with a frequency chirped signal with a linear frequency sweep.\ 
The initial and end frequencies were chosen such that the range of interest shown in ~\cref{fig:BeamResponse8thLowerHarm,fig:BeamResponse9thUpperHarm} are included in the sweep.\
This frequency range corresponded to a bandwidth of $\Delta q_{\text{BW}} = 0.026$ in tune space and thus $q_{\text{exc}} \in [n \pm q_0 - \Delta q_{\text{BW}}/2, n \pm q_0 + \Delta q_{\text{BW}}/2]$ with $n$ being the betatron sideband under investigation ($n \in [8,9]$).\

The sampling rate of both beam oscillation amplitude and stimulus signal is set to 20 samples per turn.\
As described in Section~\ref{sec:experiment} the sweep time of one scan is \SI{10}{\second} and is automatically set to fulfil the time  the IF filter of the VNA (\SI{70}{\hertz} to \SI{100}{\hertz}) needed to resolve the frequency component for 701 measurement points.\ 
This translates to approx.~$3.1\times 10^{4}$ turns per measurement point.\
With the available computing resources and the GPU implementation offered by Xsuite we are able to easily reach $3\times 10^{4}$ turns per measurement point under a computing time of approx.~\SI{3.5}{\hour} to \SI{6}{\hour} for each scan.\  

The simulated beam oscillation amplitude is then analysed together with the input stimulus signal.\ 
The magnitude $A$ and phase $\theta$ of a frequency component $f$ in the oscillation amplitude signal $S_1$ with regard to the excitation signal $S_2$ is given by
\begin{align}
A &= |I_2|/|I_1|,
\label{eq:BeamResponseAmplitude}
\end{align}
\begin{align}
\theta = \arctan{\frac{\Im{I_1}}{\Re{I_1}}} - \arctan{\frac{\Im{I_2}}{\Re{I_2}}},
\label{eq:BeamResponsePhase}
\end{align}
with
\begin{align}
 I_{1,2} = \int_{t_1}^{t_2} S_{1,2}(t) e^{i2\pi ft} \dd t.
\end{align}

With help of Eqs.~\ref{eq:BeamResponseAmplitude} and \ref{eq:BeamResponsePhase} the measured curves are recovered from the simulation.\ 
In the data analysis the recovered phase given by Eq.~\ref{eq:BeamResponsePhase} was intentionally displaced to $\theta \in [-\pi,\pi]$ and then unwrapped.

\subsection{Simulation results}
\subsubsection{Upper 9th betatron band}
\begin{figure}
  \includegraphics[width=1\linewidth]{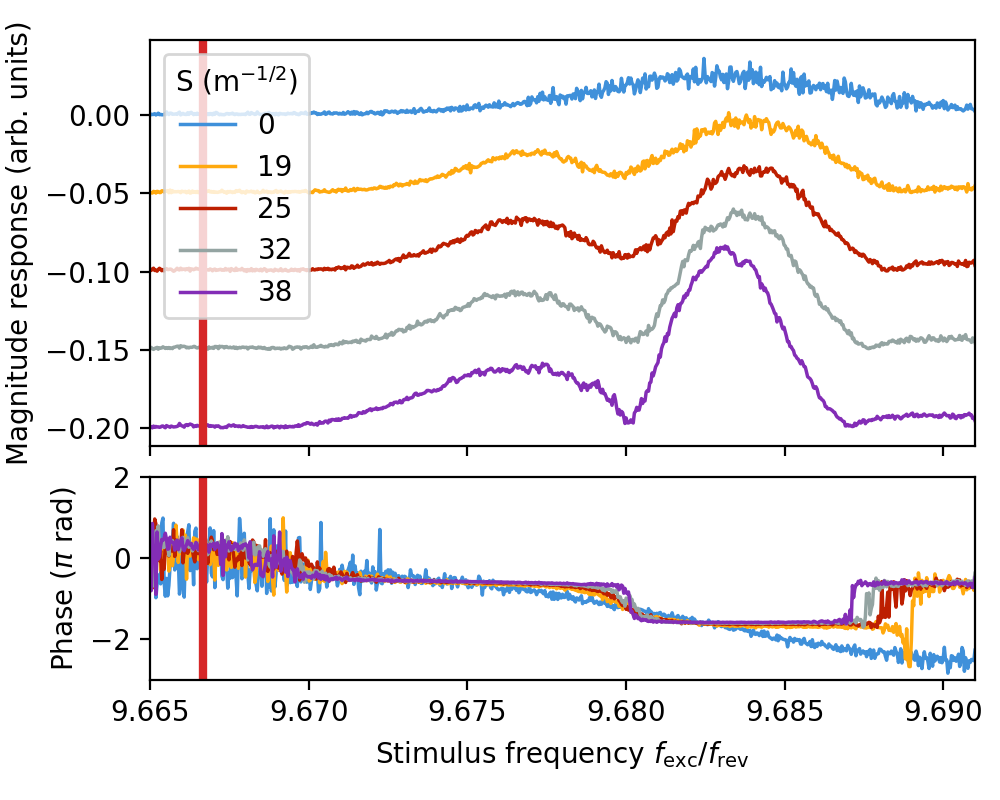}
  \caption{
  Simulated beam response of a sextupole strength scan with chromaticity corrected.\ No particle loss is observed.}
 \label{fig:SimResults9thBand}
\end{figure}

The results from the simulation are shown in Fig.~\ref{fig:SimResults9thBand}.\ The splitting behaviour is visible, although one can recognize that the threshold of $S$ at which this happens is lower than in the measurement.\ 
One important addition by the simulation is the capability to fully recover the phase, which is shown in the lower panel of Fig.~\ref{fig:SimResults9thBand}.\

The curves shown in Fig.~\ref{fig:SimResults9thBand} show a reasonable qualitative agreement with the measurements shown in Fig.~\ref{fig:BeamResponse9thUpperHarm} for $S \geq$~\SI{25}{m}$^{-1/2}$.\

The phase relation between driving excitation and centroid oscillations at the two visible dips are subsequently $-\pi$ and $\pi$. 0 and $2\pi$ degree phase represents the case where the beam centroid motion is in phase with the excitation signal while $\pi$ degrees is when they are in anti-phase. In the corrected chromaticity case shown in Fig.~\ref{fig:BeamResponse9thUpperHarm}, we note that the excitation frequency and centroid motion are in phase, then anti-phase and finally again in phase.\

\begin{figure}
  \includegraphics[width=1\linewidth]{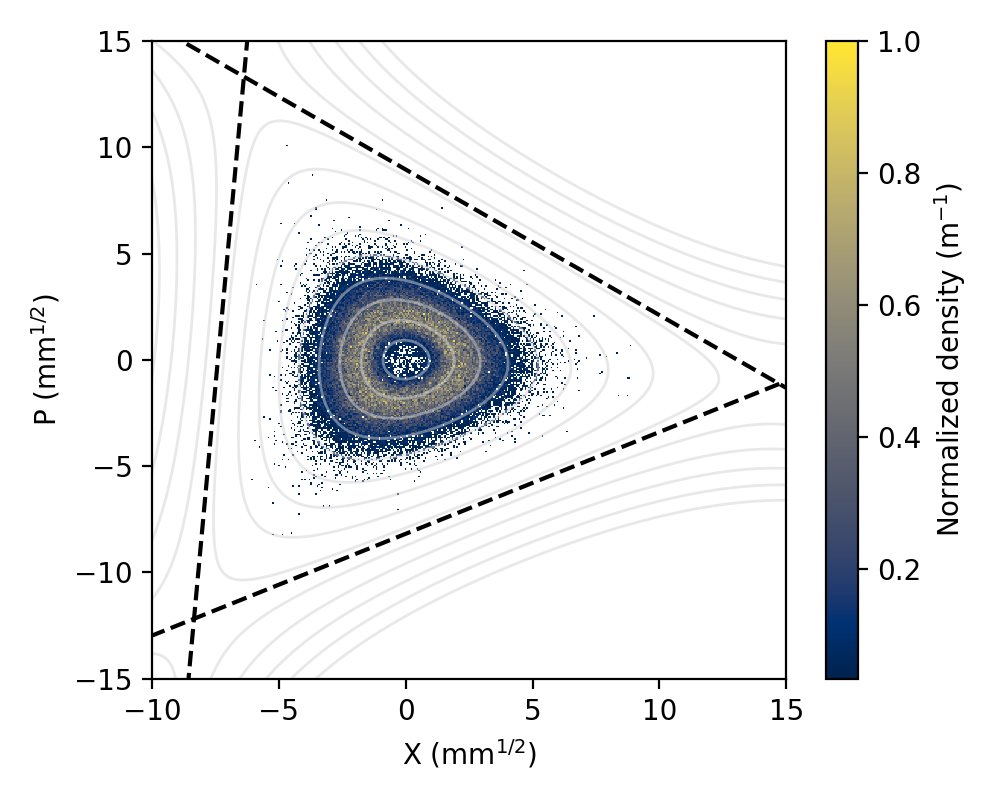}
  \caption{
  Resulting particle density distribution after simulated measurement for the 9th betatron upper band.\ The sextupole strength was $S =$~25~m$^{-1/2}$.\ The density distribution was recovered from the simulated 10$^{5}$ particles.\ The black line shows the approximate expected separatrix.}
 \label{fig:HollowBeam}
\end{figure}

\begin{figure}
  \includegraphics[width=1\linewidth]{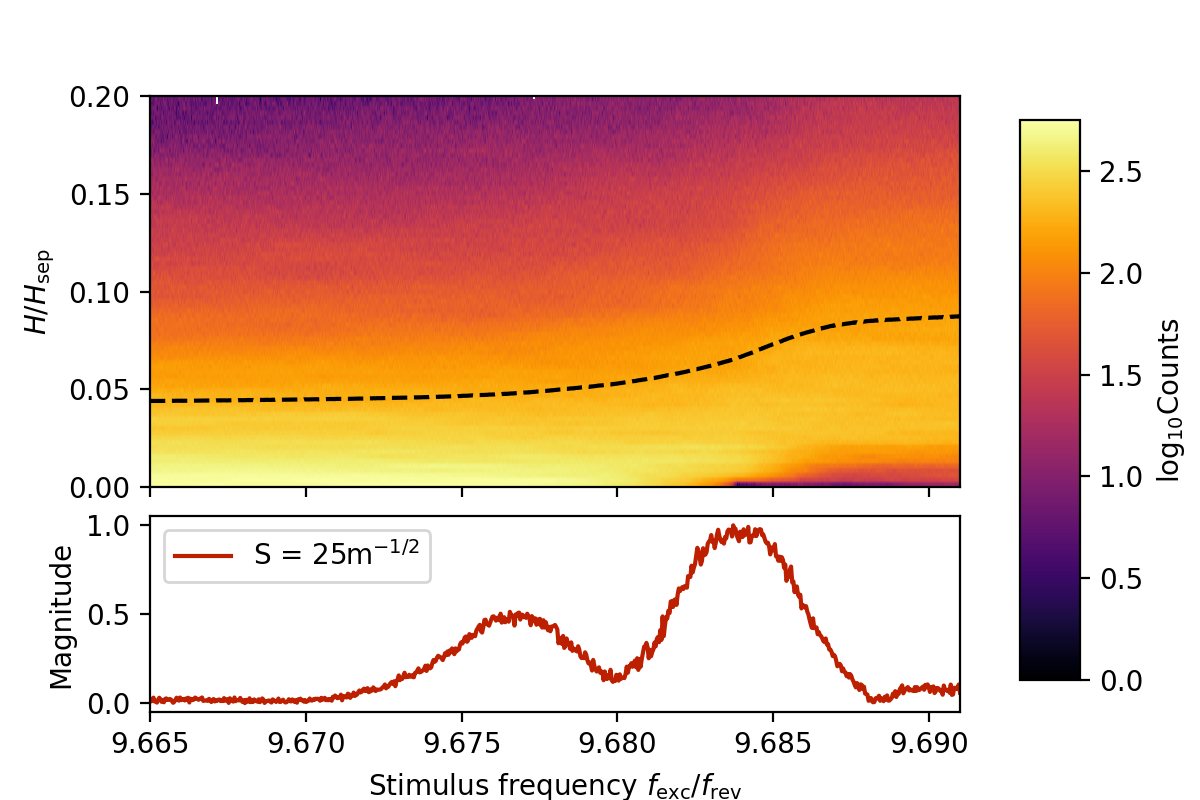}
  \caption{Effective Hamiltonian distribution as a function of the excitation frequency.\ In the lower panel for reference the simulated BTF measurement with $S =$~25~m$^{-1/2}$ is shown.\ The Hamiltonian distribution was recovered from a sample of 10$^4$ particles.}
 \label{fig:HEvolution_9thBand_S25}
\end{figure}

To understand the nature of the splitting of the measured spectra, the phase space was investigated in simulation.\ 
The resulting distribution density is shown in Fig.~\ref{fig:HollowBeam}, where it is revealed that the beam core is considerably depleted.\ 
From the simulation results we can also recover the time evolution of the particle distribution.\ 
An animation of the excitation process of the measurement is included in \textit{Supplemental Material}.\ 
To identify the moment at which the core is depleted, the evolution of the effective energy distribution of a subset of $10^{4}$ particles is shown in Fig.~\ref{fig:HEvolution_9thBand_S25}.\ One can recognize that when the excitation frequency crosses the visible dip, the average effective energy is increased and the low energy regime (core) starts to deplete.

\subsubsection{Lower 8th betatron band}

\begin{figure}
  \includegraphics[width=1\linewidth]{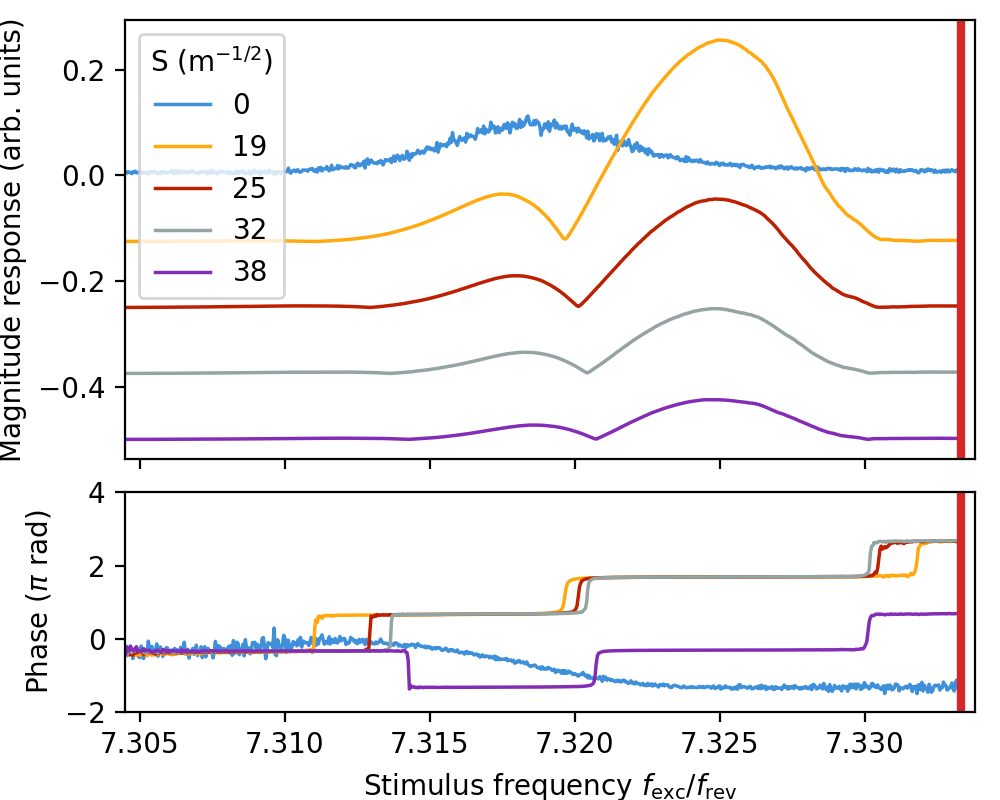}
  \caption{
  Simulated beam response of a sextupole strength scan.\ The chromaticity was fully corrected for this simulation.\ The magnitude response for the $S = 0$ was magnified by a factor 100.}
 \label{fig:SimResults8thBand}
\end{figure}

\begin{figure}
  \includegraphics[width=1\linewidth]{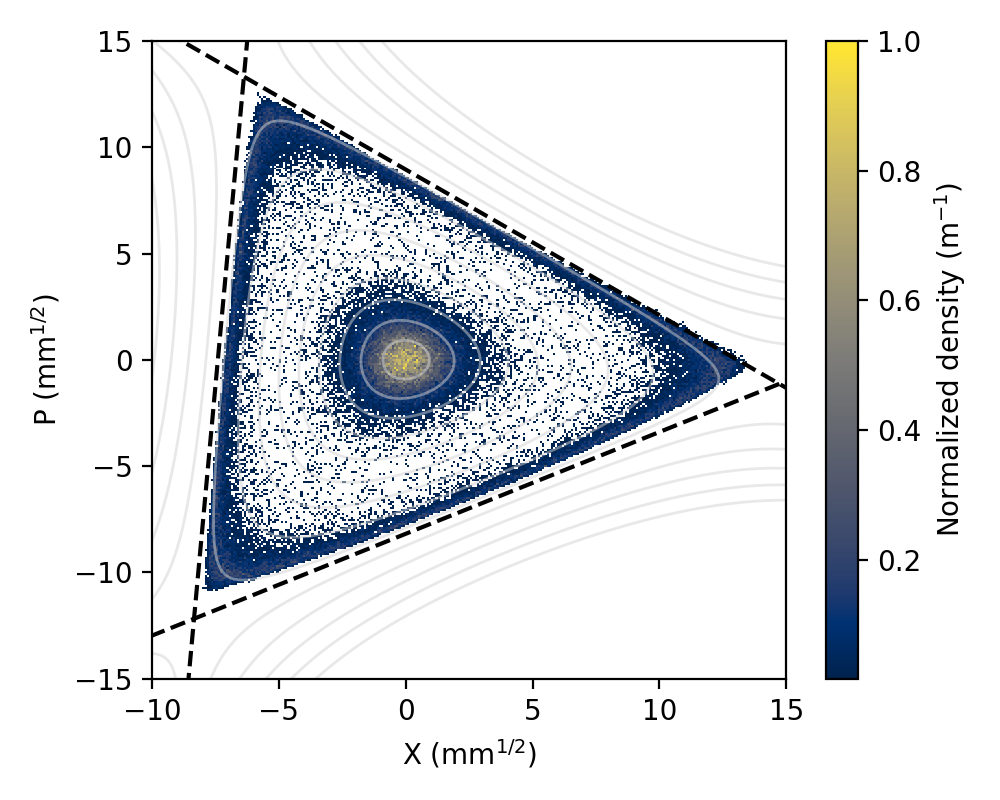}
  \caption{
  Resulting particle density distribution after simulated measurement at the lower 8th betatron sideband.\ The sextupole strength was $S =$~25~m$^{-1/2}$.\ The density distribution was recovered from the simulated 10$^{5}$ particles.\ The black line shows the approximate expected separatrix from the Kobayashi theory.}
 \label{fig:DiffusedBeam}
\end{figure}

The scan around the lower 8th betatron band was simulated as well.\ The results of the simulation are shown in Fig.~\ref{fig:SimResults8thBand} and qualitative behaviour is reproduced.\ 
The magnitude of the response splits in different regions.\
The phase of the signal shows a similar behaviour as for the previous case where the 9th upper betatron band was investigated.\
There are three visible phase jumps, one notably at the visible dip in the spectrum.\
With the results from the simulation the particle and the effective energy distributions were closely investigated.\
An animation is provided in \textit{Supplemental Material} and reveals interesting details of the evolution of the particle distribution during the excitation process.\
The resulting distribution density is illustrated in Fig.~\ref{fig:DiffusedBeam}.\
One recognizes that through the external stimulus particles diffuse towards the edge of the separatrix creating a halo.\
The simulation yields considerable particle extraction.\
The highest extraction yield starts at $S =$~19~m$^{-1/2}$ with 23\% of particle loss and decreases by approximately 2\% for each $S$ step until 17.4\%.\
For the case when the sextupoles were turned off ($S = 0$) there was no observed particle loss, as expected.\
In the measurement in contrast, there was no considerable particle loss observed.\

\begin{figure}
  \includegraphics[width=1\linewidth]{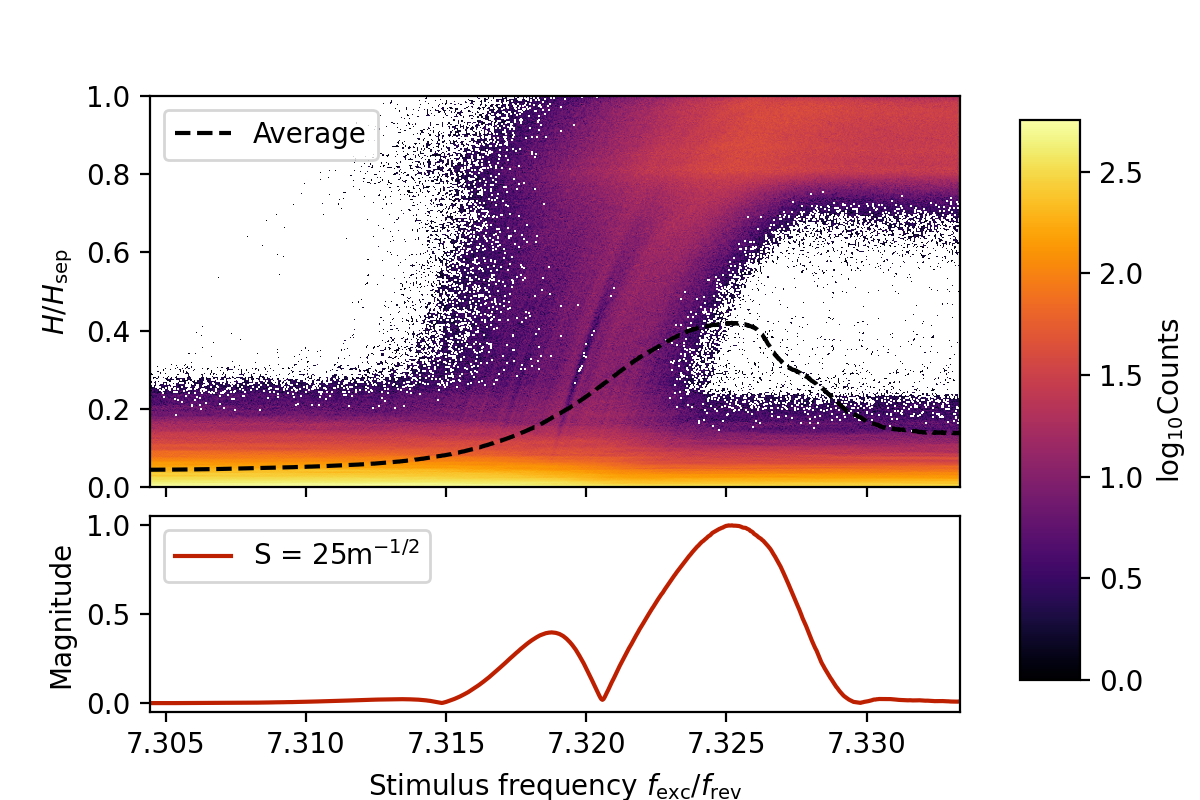}
  \caption{Effective Hamiltonian distribution as a function of the excitation frequency.\ In the lower panel for reference the simulated BTF measurement with $S =$~25~m$^{-1/2}$ is shown.\ The Hamiltonian distribution was recovered from a sample of 10$^4$ particles.}
 \label{fig:HEvolution_8thBand_S25}
\end{figure}

The creation of the halo can also be understood by investigating closer the evolution of the effective energy distribution and this is illustrated in Fig.~\ref{fig:HEvolution_8thBand_S25}.\ 
One recognizes that through the excitation process, the distribution of the energy broadens and eventually splits, leaving an empty space between core and halo (see Fig.~\ref{fig:DiffusedBeam}).\ In Fig.~\ref{fig:HEvolution_8thBand_S25} the distribution was recovered by sampling a subset of 10$^4$ particles.

A parallel simulation study with Maptrack \cite{Maptrack} was pursued.\
This study provides further analysis of single particle frequency behaviours as a link to the emergent collective dynamics beam response.\
The results are presented in \cite{Taylor_thesis} and are complementary to the results presented in this contribution.

\section{Discussion}
\label{sec:summAndDiscussion}

The measured beam transfer function shows a very particular splitting behaviour.\ 
The qualitative behaviour for the upper 9th band could be recovered in the simulation for $S \geq$~\SI{25}{m}$^{-1/2}$.\ A direct quantitative comparison was not possible, this is because the experimental conditions are in general not completely known.\
The simulations show that there is a strong dependence on the emittance for the case where $S > 0$ and this parameter could not be directly inferred in the experimental setup.\ 
Further, a shot-to-shot variation is present and can be accentuated if the number of averaging cycles is increased.\
As can be observed, the lower panel of Fig.~\ref{fig:BeamResponse9thUpperHarm} possesses a lower resolution than the case where the natural chromaticity was left uncorrected, since the scan was performed over a larger frequency span that extends over the shown region ($f_{\text{exc}}/f_{\text{rev}} \in [9.632, 9.694]$) .\
This prohibits resolving for the details that are more evident in the upper panel.\
Nonetheless both curves confirm that the splitting behaviour in the measurement is induced by increasing the resonance driving term $S$.\

In the simulation, no collective effects such as impedance or particle-particle interactions were included, thus there are no damping mechanisms that could counteract the diffusion of the particle distribution.\
This could also in part explain why the threshold of $S$, where the splitting happens appears to be higher in the measurement than the simulation.\
One other possible explanation could be that the calculated driving term $S$ is lower than calculated from set parameter values in the machine.\
The impact of intensity dependent effects is expected to be negligible, since the particle number is very low (see Table \ref{tab:ExpSetup}).\

With these considerations in mind, it appears reasonable to interpret the measured splitting behaviour as the direct consequence of a depletion of the beam core as shown in Fig.~\ref{fig:HollowBeam}.\ 
A direct confirmation of this interpretation will require a more involved experimental campaign, where the beam distribution density can be tomographed.\

The discussion continues with the results obtained for the lower 8th betatron band.\ 
The underlying dynamics is similar than for the other investigated betatron band and thus the qualitative behaviour is similar.\ 
The measured BTF appears to be split more prominently multiple times.\
The measurement was performed with a higher excitation strength.\ 
The simulation and measurement agree modestly, but only a single splitting can be reproduced.\
The simulation shows that particles acquire energy as the stimulus sweeps and comes closer to the expected region where the tune distribution should be.\
The energy gain happens even before the stimulus hits the tune value as given from the linear machine.\
This prompt diffusion combined with the amplitude detuning that is described by Eq.~\ref{eq:PNiedermayerAveDetuning} leads to a further gain in energy of particles that are in a region more closer to the separatrix.\ 
A considerable amount of particles were lost, whereas in the measurement there was no particle loss.\
Again one can argue that some collective effects that would ease the introduction of damping mechanisms are not considered in the simulation.\
The reason of this strong difference between simulation and measurement remains unclear.\
The interpretation of the results can be as well verified with a dedicated experimental campaign, where the particle distribution is tomographed to confirm if a beam core with considerable halo is generated.\
This is beyond the scope of this current study.\
\subsection{Summary}

In this contribution we presented the beam response of a coasting beam with a horizontal tune lying near the third order resonance.\ 
The resonance was driven actively by exciting the resonance driving term with strong sextupoles.\ 
To the best knowledge of the authors, the beam transfer function near an actively driven resonance has been scarcely investigated.\
These conditions are for instance, typical for resonant slow extraction from synchrotrons.\
The single particle dynamics was discussed and the average detuning due to the phase-amplitude term in the Kobayashi Hamiltonian was evaluated.\ 
The results of the semi-analytical approach agree well with the tracking results.\
Further, the beam transfer function under these conditions was measured, where a scan of the sextupole strength, the excitation strength and the excitation sweep direction were presented.\ All the measurements were performed at HIT with a Carbon-ion beam $^{12}\text{C}^{6+}$ with $E_{\text{kin}} = \SI{124.25}{MeV/nucleon}$.\
The results show a clear splitting in the measured signal and its behaviour is strongly dictated by the sextupole strength, the sweep direction and the excitation strength.\
The sextupole strength scan measurements were simulated with a realistic number of turns and excitation strength but otherwise with scaled parameters.\ 
The measurement and the simulation show a good qualitative agreement for the case of the 9th upper betatron band and some agreement for the 8th lower band.\
A high particle loss was observed in the simulation for the 8th lower band, whereas in the measurement none was recorded.\
The simulation results also reveal some hints for the interpretation of the measurements.\ 
The particle distribution under these conditions appear to ease the intake of energy, which finally distorts the distribution either by creating a hollow beam or by creating a halo at the edge of the separatrix.\

\section{Conclusion}
In conclusion, the study of the beam transfer function with a beam under resonant slow extraction conditions shows rich dynamics, which are too numerous to be discussed in a single contribution.\
For instance, the beam transfer function can be retrieved with other type of stimuli such as noise or delta excitations.\
The sweep excitation method revealed that the intake of energy of the beam is non-negligible under these conditions.\  
As an outlook, if our interpretation is correct, the sweep method can offer a way to generate hollow beams or ways to populate the halo with relatively low effort.\ 

\begin{acknowledgments}
We would like to thank the whole HIT and GSI operation teams for their warm support in the measurement campaings.\ We would also like to kindly thank P.~Forck for his support on this project.\
This project has received funding from the European Union’s Horizon 2020 Research and Innovation programme under GA No 101004730.\
This research was supported in part through the Maxwell computational resources operated at Deutsches Elektronen-Synchrotron DESY, Hamburg, Germany.
\end{acknowledgments}

\bibliography{myBibFile.bib}

\appendix

\section{Resonance driving term}
\label{sec:ResonanceDrivingTerm}
The resonance driving term $S$ can be interpreted as the normalized sextupole strength of an effective single virtual sextupole in the lattice.\
The virtual sextupole strength $S$ and its phase advance $\mu_S$ are given by
\begin{align}
    S &= |\sum_n^{N} S_n e^{-i3\mu_n}|, & S_n = \frac{1}{2}\int \beta_{x}^{3/2}k'_{S}\dd s,
    \label{eq:VirtSextupoleMag}
\end{align}
\begin{align}
    \tan{3\mu_S} &= \sum_n^{N} S_n \sin{3\mu_n}/\sum_n^{N} S_n \cos{3\mu_n}.
\end{align}
where $k_S' = \frac{1}{B\rho}\frac{\partial^2 B_y}{\partial x^2}$ is the normalized sextupole strength, $L_s$ the effective length of the sextupole and $\beta_x$ the horizontal beta-function in the Courant-Snyder parametrization $(\beta, \alpha, \gamma)$ of the horizontal phase space.

\end{document}